\PassOptionsToPackage{unicode}{hyperref}
\PassOptionsToPackage{hyphens}{url}
\PassOptionsToPackage{dvipsnames,svgnames,x11names}{xcolor}
\documentclass[
  12pt]{article}

\usepackage{amsmath,amssymb}
\usepackage{iftex}
\ifPDFTeX
  \usepackage[T1]{fontenc}
  \usepackage[utf8]{inputenc}
  \usepackage{textcomp} 
\else 
  \usepackage{unicode-math}
  \defaultfontfeatures{Scale=MatchLowercase}
  \defaultfontfeatures[\rmfamily]{Ligatures=TeX,Scale=1}
\fi
\usepackage{lmodern}
\ifPDFTeX\else  
\fi
\IfFileExists{upquote.sty}{\usepackage{upquote}}{}
\IfFileExists{microtype.sty}{
  \usepackage[]{microtype}
  \UseMicrotypeSet[protrusion]{basicmath} 
}{}
\makeatletter
\@ifundefined{KOMAClassName}{
  \IfFileExists{parskip.sty}{%
    \usepackage{parskip}
  }{
    \setlength{\parindent}{0pt}
    \setlength{\parskip}{6pt plus 2pt minus 1pt}}
}{
  \KOMAoptions{parskip=half}}
\makeatother
\usepackage{xcolor}
\makeatletter
\ifx\paragraph\undefined\else
  \let\oldparagraph\paragraph
  \renewcommand{\paragraph}{
    \@ifstar
      \xxxParagraphStar
      \xxxParagraphNoStar
  }
  \newcommand{\xxxParagraphStar}[1]{\oldparagraph*{#1}\mbox{}}
  \newcommand{\xxxParagraphNoStar}[1]{\oldparagraph{#1}\mbox{}}
\fi
\ifx\subparagraph\undefined\else
  \let\oldsubparagraph\subparagraph
  \renewcommand{\subparagraph}{
    \@ifstar
      \xxxSubParagraphStar
      \xxxSubParagraphNoStar
  }
  \newcommand{\xxxSubParagraphStar}[1]{\oldsubparagraph*{#1}\mbox{}}
  \newcommand{\xxxSubParagraphNoStar}[1]{\oldsubparagraph{#1}\mbox{}}
\fi
\makeatother

\usepackage{longtable,booktabs,array}
\usepackage{calc} 
\usepackage{etoolbox}
\makeatletter
\patchcmd\longtable{\par}{\if@noskipsec\mbox{}\fi\par}{}{}
\makeatother
\IfFileExists{footnotehyper.sty}{\usepackage{footnotehyper}}{\usepackage{footnote}}
\makesavenoteenv{longtable}
\usepackage{graphicx}
\makeatletter
\def\maxwidth{\ifdim\Gin@nat@width>\linewidth\linewidth\else\Gin@nat@width\fi}
\def\maxheight{\ifdim\Gin@nat@height>\textheight\textheight\else\Gin@nat@height\fi}
\makeatother
\setkeys{Gin}{width=\maxwidth,height=\maxheight,keepaspectratio}
\makeatletter
\def\fps@figure{htbp}
\makeatother

\makeatletter
\@ifpackageloaded{caption}{}{\usepackage{caption}}
\AtBeginDocument{%
\ifdefined\contentsname
  \renewcommand*\contentsname{Table of contents}
\else
  \newcommand\contentsname{Table of contents}
\fi
\ifdefined\listfigurename
  \renewcommand*\listfigurename{List of Figures}
\else
  \newcommand\listfigurename{List of Figures}
\fi
\ifdefined\listtablename
  \renewcommand*\listtablename{List of Tables}
\else
  \newcommand\listtablename{List of Tables}
\fi
\ifdefined\figurename
  \renewcommand*\figurename{Figure}
\else
  \newcommand\figurename{Figure}
\fi
\ifdefined\tablename
  \renewcommand*\tablename{Table}
\else
  \newcommand\tablename{Table}
\fi
}
\@ifpackageloaded{float}{}{\usepackage{float}}
\floatstyle{ruled}
\@ifundefined{c@chapter}{\newfloat{codelisting}{h}{lop}}{\newfloat{codelisting}{h}{lop}[chapter]}
\floatname{codelisting}{Listing}

\makeatother
\makeatletter
\@ifpackageloaded{caption}{}{\usepackage{caption}}
\@ifpackageloaded{subcaption}{}{\usepackage{subcaption}}
\makeatother

\ifLuaTeX
  \usepackage{selnolig}  
\fi
\usepackage[]{natbib}
\usepackage{bookmark}

\IfFileExists{xurl.sty}{\usepackage{xurl}}{} 
\hypersetup{
  pdftitle={Title},
  pdfauthor={Author 1; Author 2},
  pdfkeywords={3 to 6 keywords, that do not appear in the title},
  colorlinks=true,
  linkcolor={blue},
  filecolor={Maroon},
  citecolor={Blue},
  urlcolor={Blue},
  pdfcreator={LaTeX via pandoc}}

\newcommand{\anon}{1}

\usepackage{mathrsfs}
\usepackage{natbib, graphicx, color, setspace, amssymb, amsmath, amsthm, amstext, booktabs, fancyhdr, multirow, float, bm, lineno, xr, enumerate, comment} 
\usepackage{times}
\allowdisplaybreaks 
\usepackage{pdflscape}
\usepackage{placeins} 
\usepackage{authblk} 
\usepackage{longtable}
\usepackage[ruled,vlined,onelanguage]{algorithm2e}

\usepackage{xcite}

\usepackage{xr}
\makeatletter
\newcommand*{\addFileDependency}[1]{
  \typeout{(#1)}
  \@addtofilelist{#1}
  \IfFileExists{#1}{}{\typeout{No file #1.}}
}
\makeatother

\newcommand*{\myexternaldocument}[1]{%
    \externaldocument{#1}%
    \addFileDependency{#1.tex}%
    \addFileDependency{#1.aux}%
}

\myexternaldocument{supp_PREB}

\newcommand*\patchAmsMathEnvironmentForLineno[1]{%
  \expandafter\let\csname old#1\expandafter\endcsname\csname #1\endcsname
  \expandafter\let\csname oldend#1\expandafter\endcsname\csname end#1\endcsname
  \renewenvironment{#1}%
     {\linenomath\csname old#1\endcsname}%
     {\csname oldend#1\endcsname\endlinenomath}}%
\newcommand*\patchBothAmsMathEnvironmentsForLineno[1]{%
  \patchAmsMathEnvironmentForLineno{#1}%
  \patchAmsMathEnvironmentForLineno{#1*}}%
\AtBeginDocument{%
\patchBothAmsMathEnvironmentsForLineno{equation}%
\patchBothAmsMathEnvironmentsForLineno{align}%
\patchBothAmsMathEnvironmentsForLineno{flalign}%
\patchBothAmsMathEnvironmentsForLineno{alignat}%
\patchBothAmsMathEnvironmentsForLineno{gather}%
\patchBothAmsMathEnvironmentsForLineno{multline}%
}

\usepackage{etoolbox}
\usepackage{tikz}
\usepackage{array}

\def\bmbeta{\bm{\beta}}
\def\bmtheta{\bm{\theta}}
\def\bmSigma{\bm{\Sigma}}

\def\lsk{\left(}
\def\rsk{\right)}
\def\lbk{\left \{ }
\def\rbk{\right \} }

\newrobustcmd*{\mysquare}[1]{\tikz{\filldraw[draw=#1,fill=#1] (0,0)
		rectangle (0.2cm,0.2cm);}}

\newrobustcmd*{\mycircle}[1]{\tikz{\filldraw[draw=#1,fill=#1] (0,0) circle [radius=0.1cm];}}

\newrobustcmd*{\mytriangle}[1]{\tikz{\filldraw[draw=#1,fill=#1] (0,0) --
		(0.2cm,0) -- (0.1cm,0.2cm);}}

\usepackage{hyperref}
\hypersetup{colorlinks = true, linkcolor = blue, urlcolor = blue, citecolor = blue}


\newcounter{parentnumber}

\begin{document}

\def\spacingset#1{\renewcommand{\baselinestretch}%
{#1}\small\normalsize} \spacingset{1}


\if1\anon
{
  \title{\bf A Proportional Random Effect Block Bootstrap for General Clustered Data}
  \author{Zhi Yang Tho\thanks{Corresponding author: ZhiYang.Tho@anu.edu.au; Research School of Finance, Actuarial Studies and Statistics, The Australian National University, Canberra, ACT 2600, Australia.}} 
	\author{Raymond Chambers}
    \author{A.H. Welsh} 
	\affil{Research School of Finance, Actuarial Studies and Statistics, The Australian National University, Australia}
    \date{}
  \maketitle
} \fi

\if0\anon
{
  \bigskip
  \bigskip
  \bigskip
  \begin{center}
    {\LARGE\bf A Proportional Random Effect Block Bootstrap for General Clustered Data}
\end{center}
  \medskip
} \fi

\bigskip
\begin{abstract}
Clustered data arise naturally in many scientific and applied research settings where units are grouped within clusters. Such data are commonly analyzed using linear mixed models to account for within-cluster correlations. This article proposes a proportional random effect block bootstrap applicable to general linear mixed model settings with imbalanced cluster sizes, both random intercepts and random slopes, and autocorrelation within clusters, while allowing for non-normal random effect and error distributions. It generalizes the original random effect block bootstrap, which was developed for more restrictive settings with balanced cluster sizes, random intercepts only, and constant within-cluster correlation. The proposed bootstrap is shown to be Fisher consistent under these more general settings. Simulations demonstrate strong finite sample inferential performance relative to the original random effect block bootstrap and several existing bootstrap methods for clustered data across a variety of scenarios. Application to the Mayo Clinic primary biliary cirrhosis dataset, which contains cluster sizes ranging from 1 to 16 and exhibits evidence of within-cluster autocorrelation and non-normality, further illustrates improved bootstrap confidence intervals using the proposed method.
\end{abstract}

\noindent%
\vfill

\newpage
\spacingset{1.8} 

\section{Introduction}\label{sec:intro}

The bootstrap was introduced by \cite{efron1979} as a method to estimate the sampling distribution of a statistic using independent and identically distributed (i.i.d.) data. It has since been extended to dependent data, such as time series \citep{hans1989, bulhmann1997,dimitriosETAL2019, friedrichANDlin2024} and spatial data \citep{lahiriANDzhu2006, sergioETAL2019}. This article focuses on bootstrap methods for a specific form of dependent data, namely clustered data, in which observations are grouped into known clusters. Clustered data commonly arise in various fields, such as healthcare where repeated visits are nested within patients, education where students are nested within schools, and economics where firms are nested within industries. A widely used framework for analyzing such data is the linear mixed model \citep[LMM,][]{pinheiroANDbates2000,batesETAL2015}, which incorporates fixed effects to capture population-level effects and random effects to account for within-cluster dependence and between-cluster variability.

We propose a proportional random effect block  (PREB) bootstrap that accommodates general clustered data settings, including imbalanced cluster sizes, models with both random intercepts and random slopes,  and within-cluster autocorrelation, as well as non-normal random effect and error distributions. The method is motivated by increasingly complex clustered data structures arising from modern data collection processes. For example, the Mayo Clinic primary biliary cirrhosis dataset (Section \ref{sec:realdata}) is a longitudinal data consisting of repeated visits (units) nested within patients (clusters), with cluster sizes ranging from 1 to 16, while the continuous biochemical markers for each patient are not only non-normal but also exhibit serial dependence across visits. These features highlight the need for bootstrap procedures that remain valid under such complex clustered data structure.

The PREB bootstrap is a generalization of the semiparametric random effect block (REB) bootstrap studied by \cite{chambersANDchandra2013}, which was originally developed to handle non-normal random effect and error distributions in more restrictive clustered data settings; namely, balanced cluster sizes and random-intercept-only models with constant correlation within clusters. Like the REB bootstrap, the PREB procedure resamples cluster-level predicted random effects,  samples clusters (``blocks''), and resamples residuals within the sampled clusters. To accommodate imbalanced cluster sizes, the method applies appropriate centering and rescaling to the predicted random effects and residuals prior to resampling, and replaces simple random sampling in the REB bootstrap with probability-proportional-to-size (PPS) sampling for cluster selection. Extension to incorporate random slopes is achieved by scaling the predicted random effect vectors using the Cholesky decomposition of their estimated covariance matrix. Finally, to allow for within-cluster autocorrelation, residuals are first transformed into uncorrelated components before resampling and are then reconstructed to preserve an autocorrelated structure within each cluster. To the best of our knowledge, this article is among the first studies to propose a random effect bootstrap method that can be applied to such general clustered data.

Both the PREB and REB bootstraps belong to a broader class of random effect or residual bootstraps, which involve resampling predicted random effects and/or residuals. Other methods within this class include the parametric bootstrap \citep{butarANDlahiri2003, kubokawaANDnagashima2012} which samples random effects and residuals from normal distributions with the corresponding estimated variance components. However, the validity of the parametric bootstrap depends heavily on the stochastic assumption of the model e.g., the normality assumptions of the random effects and residuals. Semiparametric versions of the random effect or residual bootstraps have therefore been developed to circumvent this limitation of the parametric bootstrap \citep[see e.g.,][]{carpenterETAL2003,relugaETAL2024}, including the REB bootstrap discussed above. Another class of bootstrap methods for clustered data, not directly motivated by LMMs, has also been studied in parallel. One example is the cluster bootstrap \citep{davisonANDhinkley1997, mccullagh2000}, in which clusters are randomly sampled with replacement, optionally followed by random permutation or resampling of observations within clusters. \cite{fieldETAL2010} and \cite{oshaughnessyANDwelsh2018} studied the generalized cluster bootstrap for clustered data, which involves resampling weights associated with the estimating equations. 

While various bootstrap methods has been proposed for clustered data as reviewed above, there have been only limited studies investigating their performance under general settings where cluster sizes may be imbalanced,  random slopes may be required in addition to random intercepts, observations within clusters may exhibit autocorrelation, and/or random effects and errors may deviate from normality. A related study by \cite{samantaANDwelsh2013} showed that the generalized cluster bootstrap outperformed the transformation bootstrap under imbalanced cluster sizes, although their study considered only random-intercept models with constant within-cluster correlation. Given the importance of bootstraps in conducting inferences \citep{efronANDtibshirani1994, davisonANDhinkley1997}, this article aims to bridge this gap in the literature by studying the PREB bootstrap, which is designed to provide satisfactory confidence interval coverage for parameters in LMMs under general clustered data settings.
 
We demonstrate theoretically that the proposed PREB bootstrap is Fisher consistent even when cluster sizes are imbalanced and the residuals exhibit autocorrelation. Moreover, this only requires the specification of the first two moments of the random effects and residuals; consequently, normality is not assumed. Simulation studies demonstrate empirically that the PREB bootstrap confidence intervals consistently achieve good coverage across a wide range of settings, and that they outperform the REB bootstrap and other competing bootstrap methods. An application to the Mayo Clinic primary biliary cirrhosis data further reinforces the value of the PREB bootstrap in accounting for autocorrelation across succesive visits within each patient, as well as the imbalanced number of visits across patients. In particular, the resulting confidence intervals not only show strong evidence of within-cluster autocorrelation but also differ substantially from those obtained under methods that assume constant within-cluster correlation and/or an equal number of visits within patients.

The rest of this article is organized as follows. Section \ref{sec:model} introduces the LMM and the proposed PREB bootstrap. Section \ref{sec:theory} presents theoretical results for the proposed bootstrap under general clustered data settings.
Section \ref{sec:simulation} presents results of a simulation study, while an application to the Mayo Clinic primary biliary cirrhosis data is provided in Section \ref{sec:realdata}. Section \ref{sec:conclusion} offers some concluding remarks.

\section{Linear Mixed Model and Proportional REB Bootstrap} \label{sec:model}
Let $y_{ij}$ denote the response, $\bm{x}_{ij} = (1, x_{ij,1}, \cdots, x_{ij,p-1})^\top$ denote the $p$-dimensional fixed-effect covariate vector, and $\bm{z}_{ij} = (1, z_{ij,1}, \cdots, z_{ij,q-1})^\top$ denote the $q$-dimensional random-effect covariate vector for unit $j$ in cluster $i$ for $j=1,\cdots,n_i$ and $i=1,\cdots,D$. 
There are $D$ clusters with (potentially imbalanced) cluster sizes $n_i$, and $N = \sum_{i=1}^{D} n_i$ total observations. 
Clustered data of this form are commonly modeled with linear mixed models (LMMs) to account for both between- and within-cluster variation; specifically, 
\begin{equation}
    y_{ij} = \bm{x}_{ij}^\top \bmbeta + \bm{z}_{ij}^\top \bm{u}_i + e_{ij}, 
    \textrm{for } j=1,\cdots,n_i, \; i=1,\cdots,D,
    \label{eq:lmm}
\end{equation}
where $\bmbeta = (\beta_{0},\cdots, \beta_{p-1})^\top$ is the vector of fixed effects, and $\bm{u}_i = (u_{i0}, \cdots, u_{i,q-1})^\top$ are i.i.d. $q$-dimensional cluster-level random effects with mean $\bm{0}_q$ and $q \times q$ covariance matrix $\bm{G}$. The error terms $e_{ij}$ are assumed to be independent of the random effects and to have mean zero, with a possible first-order autoregressive (AR-1) covariance structure within each cluster; that is, $\mathrm{cov}(e_{ij}, e_{ij'}) = \sigma^2_e \rho ^{|j - j'|}$, $|\rho| < 1$ is the AR-1 correlation parameter. Error terms from different clusters are assumed to be independent. The AR-1 error process can equivalently be represented as
\begin{equation}
    e_{ij} = \rho e_{i,j-1} + w_{ij} \textrm{ for } j=2,\cdots,n_i, \; \textrm{with } e_{i1} \sim (0,\sigma^2_e)  \textrm{ for } i=1,\cdots, D,
    \label{eq:ar1_error_process}
\end{equation}
where $w_{ij}$ are i.i.d. white-noise terms with mean zero and variance $\sigma^2_e (1-\rho^2)$. Importantly, our bootstrap procedure only requires specification of the first two moments for $\bm{u}_i$, $e_{ij}$ and $w_{ij}$, and therefore does not require any distributional assumptions such as normality. By allowing for an AR-1 error process, the model is particularly useful in settings where observations within clusters consist of repeated measurements over successive time points, which is common in longitudinal data analysis.

Model \eqref{eq:lmm} can be written in an equivalent vector form as
\begin{equation*}
    \bm{y} = \bm{X} \bmbeta + \bm{Z} \bm{u} + \bm{e},
\end{equation*}
where $\bm{y} = ( \bm{y}_1^\top ,\cdots, \bm{y}_D^\top )^\top$ is the $N$-dimensional vector of stacked responses, $\bm{y}_i = (y_{i1},\cdots, y_{in_i})^\top$, $\bm{X} = (\bm{X}_1^\top,\cdots, \bm{X}_D^\top)^\top$ is the $N \times p$ fixed effect model matrix, $\bm{X}_i = (\bm{x}_{i1}, \cdots, \bm{x}_{in_i})^\top$, and $\bm{Z} = \mathrm{diag}(\bm{Z}_{1},\cdots,\bm{Z}_{D})$ is the $N \times qD$ block diagonal random effect model matrix with block $i$ given by $\bm{Z}_i = (\bm{z}_{i1}, \cdots, \bm{z}_{in_i})^\top$. In addition, $\bm{u} = (\bm{u}_1^\top,\cdots,\bm{u}_D^\top)^\top $ is the $qD$-dimensional vector of stacked random effects, and $\bm{e} = (\bm{e}_1^\top, \cdots, \bm{e}_D^\top)^\top$ is the $N$-dimensional vector of stacked errors with $\bm{e}_i = (e_{i1},\cdots,e_{in_i})^\top$. 
This leads to $\mathrm{E}(\bm{y}) = \bm{X} \bmbeta$ and $\mathrm{var}(\bm{y}) = \bmSigma(\bm{G},\sigma^2_e, \rho) = \bm{Z} (\bm{I}_D \otimes \bm{G}) \bm{Z}^\top + \sigma^2_e \mathrm{diag}(\bmSigma_{1}(\rho), \cdots, \bmSigma_{D}(\rho)   )$, where $\otimes$ is the Kronecker product operator and $\bmSigma_i(\rho)$ is the $n_i \times n_i$ AR-1 correlation matrix with $(j,j')$-th entry given by $\rho^{|j-j'|}$. The fixed effects $\bmbeta$ capture the systematic impact of covariates on the response.  The random intercepts $u_{i0}$ represent cluster-specific deviations from the population-level intercept, whereas the random slopes $u_{ik}$ for $k\geq1$ capture heterogeneity in covariate effects across clusters. 

The above setting includes several special cases commonly considered in the clustered data bootstrap literature reviewed in Section \ref{sec:intro}. For example, the balanced cluster sizes setting in \cite{chambersANDchandra2013} is obtained when $n_i = n$ for all $i=1,\cdots,D$, while the random-intercept-only model considered by \cite{samantaANDwelsh2013} is recovered by setting $q = 1$. It is also common to assume independent error terms within clusters \citep[e.g., ][]{carpenterETAL2003}, which corresponds to the special case of $\rho = 0$. Finally, the Gaussian random effects and error terms assumed in \cite{butarANDlahiri2003} are also covered as a special case, since our framework only requires specification of their first two moments rather than full distributions.

\subsection{Proportional REB Bootstrap} \label{subsec:PREB1}
Let $\bmtheta = (\bmbeta^\top, \mathrm{vech}(\bm{G}), \sigma^2_e, \rho)^\top$ denote the parameter vector. The corresponding estimators $\hat{\bmtheta} = (\hat{\bmbeta}^\top, \mathrm{vech}(\hat{\bm{G}}), \hat{\sigma}^2_e, \hat{\rho})^\top$ are typically obtained via quasi-maximum likelihood (ML) or quasi-restricted maximum likelihood (REML) \citep{patternsonANDthompson1971,harville1977,batesETAL2015}; that is, by maximizing a \textit{Gaussian quasi-likelihood} even when the response distribution is not truly Gaussian. For example, the quasi-ML estimator is defined as the maximizer of 
\begin{equation*}
    l(\bmtheta; \bm{y}, \bm{X})  = \frac{1}{2} \log |\bmSigma^{-1}(\bm{G},\sigma^2_e, \rho)| - \frac{1}{2} (\bm{y} - \bm{X} \bmbeta)^\top \bmSigma^{-1}(\bm{G},\sigma^2_e, \rho) (\bm{y} - \bm{X} \bmbeta).
\end{equation*} 
Marginal residuals $r_{ij} = y_{ij} - \bm{x}_{ij}^\top \hat{\bmbeta}$ can then be computed and used to obtain the cluster-level predicted random effects and unit-level residuals as
\begin{equation}
\hat{\bm{u}}_i = (\bm{Z}_i^\top \bm{Z}_i)^{-1} \bm{Z}_i^\top \bm{r}_i, \; \hat{e}_{ij} = r_{ij} - \bm{z}_{ij}^\top \hat{\bm{u}}_i,\,\,\, j=1,\cdots,n_i, \, i=1,\cdots,D, 
\label{eq:hatu_hate}
\end{equation}
where $\bm{r}_i = (r_{i1}, \cdots, r_{in_i})^\top$. The inversion of the matrices $\bm{Z}_i^\top \bm{Z}_i$ in \eqref{eq:hatu_hate} requires that the $n_i \times q$ matrices $\bm{Z}_i$ have full column rank, which requires $n_i \geq q$ for $i=1,\cdots,D$. Moreover, to account for the potential AR-1 structure in the error terms in \eqref{eq:ar1_error_process}, we further construct decorrelated residuals
\begin{equation}
\hat{w}_{i1} = \hat{e}_{i1}, \; \hat{w}_{ij} = \hat{e}_{ij} - \hat{\rho} \hat{e}_{i,j-1},\,\,\, j=2,\cdots, n_i, \,i=1,\cdots,D.
\label{eq:hatepsilon}
\end{equation}
 These quantities, $\hat{\bm{u}}_i$,  $\hat{e}_{ij}$ and $\hat{w}_{ij}$, form the basis of our bootstrap procedure;  $\hat{\bm{u}}_i$ is a rescaled (unshrunk) version of the usual empirical best linear unbiased predictor (EBLUP). 

Specifically, the proportional random effect block (PREB) bootstrap employs the idea of ``reflating'' introduced by \cite{carpenterETAL2003} to ``unshrink'' EBLUPs, by centering and scaling the cluster-level random effects $\hat{\bm{u}}_i$ and unit-level residuals $\hat{w}_{ij}$. Define
\begin{equation*}
\hat{\bm{u}}_i^c = \hat{\bm{u}}_i - \frac{1}{D} \sum_{i'=1}^{D} \hat{\bm{u}}_{i'}, \; 
\hat{w}_{ij}^c = \hat{w}_{ij} - \frac{1}{n_i} \sum_{j' = 1}^{n_i} \hat{w}_{ij'},
\end{equation*}
and construct
\begin{equation}
    \hat{\bm{u}}_{i}^{sc} = \bm{L}_{\hat{\bm{G}}} \bm{L}_{\bm{V}}^{-1} \hat{\bm{u}}_i^c, \; \hat{w}_{ij}^{sc} = \hat{w}_{ij}^c \frac{\hat{\sigma}_e \sqrt{1 - \hat{\rho}^2} }{\sqrt{N^{-1} \sum_{i' = 1}^{D} \sum_{j' = 1}^{n_{i'}} (\hat{w}_{i' j'}^c)^2  }}, 
    \label{eq:hatu_hatepsilon_PREB}
\end{equation}
for $j=1,\cdots,n_i$ and $i=1,\cdots,D$, where $\bm{L}_{\hat{\bm{G}}}$ and $\bm{L}_V$ are Cholesky factors defined by $\hat{\bm{G}} = \bm{L}_{\hat{\bm{G}}} \bm{L}_{\hat{\bm{G}}}^\top$ and $ D^{-1} \sum_{i=1}^{D} \hat{\bm{u}}_i^{c} \hat{\bm{u}}_i^{c\top} = \bm{L}_{\bm{V}} \bm{L}_{\bm{V}}^\top$, respectively.

Let $\mathrm{SRSWR}(\mathcal{A},c) $ denote $c$ independent draws from a set $\mathcal{A}  = \{a_1,\cdots,a_D\}$ using simple random sampling (SRS) with replacement, where each element $a_i$ could be a scalar or a vector, and $\mathrm{PPSWR}(\mathcal{A}, \bm{b},c)$ denote $c$ independent draws from $\mathcal{A}$ using probability-proportional-to-size (PPS) sampling with replacement with probabilities proportional to $\bm{b} = (b_1,\cdots,b_D)^\top$, i.e., the selection probability of $a_i$ is $b_i / \sum_{i'=1}^{D} b_{i'} $. The full PREB bootstrap procedure is described in Algorithm \ref{al:PREB}.

\begin{algorithm}[tb] \spacingset{1}
\caption{\footnotesize Algorithm for the PREB bootstrap.} \label{al:PREB}

\KwIn{Point estimates $\hat{\bmtheta} = (\hat{\bmbeta}^\top, \mathrm{vech}(\hat{\bm{G}}), \hat{\sigma}^2_e, \hat{\rho})^\top$.}
  \begin{enumerate}[1.]
        \item Compute marginal residuals $\bm{r}_{i} = \bm{y}_i - \bm{X}_i \hat{\bmbeta}$ for $i = 1,\cdots,D$. Obtain cluster-level predicted random effects $\hat{\bm{u}}_i$, unit-level residuals  $\hat{e}_{ij}$ from \eqref{eq:hatu_hate}, and $\hat{w}_{ij}$ from \eqref{eq:hatepsilon}.
        \item Reflate these quantities to $\hat{\bm{u}}_i^{sc}$ and $\hat{w}_{ij}^{sc}$ for $j=1,\cdots,n_i$ and $i=1,\cdots,D$ using \eqref{eq:hatu_hatepsilon_PREB}.
  \end{enumerate}
  
\For(){$b = 1,\cdots,B$}
{
\begin{enumerate}[1.]
\setcounter{enumi}{2}
    \item Draw samples of cluster-level random effects $\bm{u}_i^{*} = \mathrm{SRSWR}( \{ \hat{\bm{u}}_1^{sc},\cdots,\hat{\bm{u}}_D^{sc} \},1)$ for $i=1,\cdots,D$.

    \item For each $i=1,\cdots,D$, sample a donor cluster $d_i^{*} = \mathrm{PPSWR}( \{1,\cdots,D\}, (n_1,\cdots,n_D) ,1)$, then draw  $\bm{w}_{i}^{*} = (w_{i1}^{*},\cdots, w_{in_{i}}^{*})^\top = \mathrm{SRSWR}( \{ \hat{w}^{sc}_{d_i 1}, \cdots, \hat{w}^{sc}_{d_i n_{d_i}}\}, n_i) $, set $e_{i1}^* = w_{i1}^* / \sqrt{1 - \hat{\rho}^2}$ and reconstruct $e_{ij}^* = \hat{\rho} e_{i,j-1}^* + w_{ij}^* $ for $j=2,\cdots,n_i$.
    
    \item Form bootstrap responses as $\bm{y}^{*} = \bm{X} \hat{\bmbeta} + \bm{Z} \bm{u}^{*} + \bm{e}^{*}$, where $\bm{u}^{*} = (\bm{u}_1^{*\top}, \cdots, \bm{u}_D^{*\top})^\top$, $\bm{e}^{*} = (\bm{e}_1^{*\top}, \cdots, \bm{e}_D^{*\top})^\top$ and $\bm{e}_i^* = (e_{i1}^*, \cdots, e_{in_i}^*)^\top$.

    \item Fit model \eqref{eq:lmm} to  $(\bm{y}^{*} , \bm{X}, \bm{Z})$ and obtain $\hat{\bmtheta}^* = (\hat{\bmbeta}^{*\top}, \mathrm{vech}(\hat{\bm{G}}^*), \hat{\sigma}^{2*}_e, \hat{\rho}^*)^\top$.
\end{enumerate}
}

\KwOut{Bootstrap samples $\{\hat{\bmtheta}^{*(b)}  : b = 1,\cdots, B\}$ with superscript $(b)$ indicating the $b$-th bootstrapped estimates.}

\end{algorithm}

The bootstrap distributions for different elements of $\hat{\bmtheta}$ obtained from Algorithm \ref{al:PREB} can be used to form bootstrap confidence intervals for the corresponding elements of the parameter vector $\bmtheta$. While many methods can be used to form such intervals \citep[see e.g.,][]{efronANDtibshirani1994,davisonANDhinkley1997}, this article concentrates on the bootstrap percentile confidence intervals. Under this method, a $100(1-\alpha)\%$ confidence interval for the $k$-th element of $\bmtheta$ (denoted $\theta_k$) is constructed as $(\hat{\theta}_{k,\alpha/2}^*, \hat{\theta}_{k,1-\alpha/2}^*)$, where $\hat{\theta}_{k,\alpha/2}^*$ and $\hat{\theta}_{k,1-\alpha/2}^*$ denote the $\alpha/2$ and $1-\alpha/2$ quantiles of the bootstrap distribution for $\hat{\theta}_k$, respectively, and $\hat{\theta}_k$ denotes the $k$-th element of $\hat{\bmtheta}$ for $k=1,\cdots, p+ q(q+1)/2 + 2  $.

Let $\mathrm{E}^*(\cdot)$, $\mathrm{var}^*(\cdot)$ and $\mathrm{cov}^*(\cdot)$  denote boostrap expectation, variance and covariance operators conditional on the observed response vector $\bm{y}$. The PREB bootstrap ensures that the first two bootstrap moments matches the corresponding estimated moments:
\begin{align}
    \mathrm{E}^*(\bm{u}_i^*) &= \bm{0}_q, \; \mathrm{E}^*(e_{ij}^*) = 0, \notag \\
    \mathrm{var}^*(\bm{u}_i^*) &= \hat{\bm{G}}, \; \mathrm{cov}^*(e_{ij}^*, e_{ij'}^*) = \hat{\sigma}^2_e\hat{\rho}^{|j-j'|}. \label{eq:moment_matching}
\end{align}
Importantly, these properties underpin the consistency of the PREB bootstrap confidence intervals; see Section \ref{sec:theory} for the detailed derivations of the corresponding bootstrap moments. In particular, they are achieved through the “reflating” step in \eqref{eq:hatu_hatepsilon_PREB}, which centers and rescales the predicted random effects and decorrelated residuals to recover the estimated covariance structure. 

The PREB bootstrap is a generalization of the prescaled REB bootstrap of \cite{chambersANDchandra2013}. Under the more restrictive setting considered in \cite{chambersANDchandra2013}, namely $n_i = n$ for all $i=1,\cdots,D$, $q=1$, and $\hat{\rho} = 0$, the PREB and prescaled REB bootstraps are equivalent and both satisfy \eqref{eq:moment_matching}. However, when cluster sizes are imbalanced, random slopes are present, or residuals exhibit AR-1 dependence within clusters, only the PREB bootstrap continues to satisfy these moment-matching properties.

The semiparametric bootstrap of \cite{carpenterETAL2003} uses some similar ideas to those used in the PREB and REB bootstraps for ``reflating'' the random effects, although they used EBLUPs rather than $\hat{\bm{u}}_i$ from \eqref{eq:hatu_hate}. Their bootstrap also differs in the treatment of unit-level residuals, as they directly resample the residuals $\hat{e}_{ij}$ rather than the decorrelated residuals $\hat{w}_{ij}$. Specifically, they generate $\bm{e}^* = \mathrm{SRSWR}( \{c \hat{e}_{11}, \cdots, c \hat{e}_{Dn_D} \}, N )$, where $c$ is a rescaling factor (see Section \ref{sec:simulation} for details), thereby resampling from the pooled collection of residuals without preserving within-cluster dependence. Consequently, their bootstrap procedure is designed for independent error terms and does not accommodate autocorrelated errors such as the AR-1 structure considered here.

\section{Theoretical Properties of Proportional REB Bootstrap} \label{sec:theory}

This section discusses the theoretical properties of the proposed PREB bootstrap under model \eqref{eq:lmm} for our general clustered data setting, allowing for the possibility of imbalanced cluster sizes, the inclusion of random slopes, within-cluster AR-1 correlation, and/or non-normal random effects and error terms. As the quasi-ML and quasi-REML estimators are asymptotically equivalent when $p$ is fixed, we focus on the quasi-ML estimator $\hat{\bmtheta}$. Following the theoretical work of \cite{shaoETAL2000} and \cite{carpenterETAL2003}, bootstrap percentile confidence intervals are Fisher consistent if the bootstrap expectations of the estimating functions are zero. The quasi-ML estimating functions are 
\begin{equation}
    \frac{\partial l(\bmtheta; \bm{y})}{\partial \bmtheta} = 
    \begin{pmatrix}
        -2\bm{X}^\top \bmSigma^{-1}(\bm{G},\sigma^2_e, \rho) (\bm{y} - \bm{X} \bmbeta) \\

        \frac{1}{2} \mathrm{tr} \lbk \bmSigma(\bm{G},\sigma^2_e, \rho) \frac{\partial \bmSigma^{-1}(\bm{G},\sigma^2_e, \rho)}{\partial g_{00}} \rbk - \frac{1}{2} \mathrm{tr} \lbk (\bm{y} - \bm{X} \bmbeta) (\bm{y} - \bm{X} \bmbeta)^\top \frac{ \partial \bmSigma^{-1}(\bm{G},\sigma^2_e, \rho)}{\partial g_{00}} \rbk \\

        \frac{1}{2} \mathrm{tr} \lbk \bmSigma(\bm{G},\sigma^2_e, \rho) \frac{\partial \bmSigma^{-1}(\bm{G},\sigma^2_e, \rho)}{\partial g_{10}} \rbk - \frac{1}{2} \mathrm{tr} \lbk (\bm{y} - \bm{X} \bmbeta) (\bm{y} - \bm{X} \bmbeta)^\top \frac{ \partial \bmSigma^{-1}(\bm{G},\sigma^2_e, \rho)}{\partial g_{10}} \rbk \\

        \vdots \\

        \frac{1}{2} \mathrm{tr} \lbk \bmSigma(\bm{G},\sigma^2_e, \rho) \frac{\partial \bmSigma^{-1}(\bm{G},\sigma^2_e, \rho)}{\partial g_{q-1,q-1}} \rbk - \frac{1}{2} \mathrm{tr} \lbk (\bm{y} - \bm{X} \bmbeta) (\bm{y} - \bm{X} \bmbeta)^\top \frac{ \partial \bmSigma^{-1}(\bm{G},\sigma^2_e, \rho)}{\partial g_{q-1,q-1}} \rbk \\

        \frac{1}{2} \mathrm{tr} \lbk \bmSigma(\bm{G},\sigma^2_e, \rho) \frac{\partial \bmSigma^{-1}(\bm{G},\sigma^2_e, \rho)}{\partial \sigma^2_e} \rbk - \frac{1}{2} \mathrm{tr} \lbk (\bm{y} - \bm{X} \bmbeta) (\bm{y} - \bm{X} \bmbeta)^\top \frac{ \partial \bmSigma^{-1}(\bm{G},\sigma^2_e, \rho)}{\partial \sigma^2_e} \rbk \\

        \frac{1}{2} \mathrm{tr} \lbk \bmSigma(\bm{G},\sigma^2_e, \rho) \frac{\partial \bmSigma^{-1}(\bm{G},\sigma^2_e, \rho)}{\partial \rho} \rbk - \frac{1}{2} \mathrm{tr} \lbk (\bm{y} - \bm{X} \bmbeta) (\bm{y} - \bm{X} \bmbeta)^\top \frac{ \partial \bmSigma^{-1}(\bm{G},\sigma^2_e, \rho)}{\partial \rho} \rbk
    \end{pmatrix},
    \label{eq:score}
\end{equation}
where $g_{kk'}$ is the $(k+1,k'+1)$-th entry of $\bm{G}$ for $k,k' = 0,\cdots,q-1$. It follows that $\mathrm{E}\{\partial l(\bmtheta; \bm{y}) / \partial \bmtheta\} = \bm{0}_{p+ q(q+1)/2 +2}$ since $\mathrm{E}(\bm{y} - \bm{X} \bmbeta) = \bm{0}_{p}$ and $\mathrm{E} \{ (\bm{y} - \bm{X} \bmbeta) (\bm{y} - \bm{X} \bmbeta)^\top \} - \bmSigma(\bm{G},\sigma^2_e, \rho) = \bm{0}_{N \times N}$. Thus, $\hat{\bmtheta}$ is Fisher consistent for $\bmtheta$. Analogously, the bootstrap estimator $\hat{\bmtheta}^*$ is consistent for $\hat{\bmtheta}$ if $\mathrm{E}^*\{\partial l(\hat{\bmtheta}; \bm{y}^*) / \partial \bmtheta\} = \bm{0}_{p+ q(q+1)/2 + 2}$, recalling that $\mathrm{E}^*(\cdot)$ is the bootstrap expectation operator conditional on the observed $\bm{y}$. With $\bm{y}^* = \bm{X} \hat{\bmbeta} + \bm{Z} \bm{u}^* + \bm{e}^*$, this holds provided that 
\begin{align*}
    \mathrm{E}^*(\bm{u}_i^*) &= \bm{0}_q, \; \mathrm{E}^*(e_{ij}^*) = 0, \\
    \mathrm{E}^*(\bm{u}_i^{*} \bm{u}_i^{*\top}) &= \hat{\bm{G}}, \;  \mathrm{E}^*(e_{ij}^{*} e_{ij'}^*) = \hat{\sigma}^2_e \hat{\rho}^{|j-j'|}, 
\end{align*}
since $\bm{u}_i^*$ and $e_{ij}^*$ are bootstrapped independently. We next show that the PREB bootstrap satisfies these conditions.

Recall that the PREB bootstrap samples $\bm{u}_i^*$ from $\hat{\bm{u}}_i^{sc}$ using SRS. We then obtain
\begin{align*}
    \mathrm{E}^*(\bm{u}_i^*) &= \sum_{i' = 1}^{D} \mathbb{P}( \bm{u}_i^* = \hat{\bm{u}}_{i'}^{sc}) \times \hat{\bm{u}}_{i'}^{sc} = \sum_{i' = 1}^{D} \frac{1}{D} \times \hat{\bm{u}}_{i'}^{sc} = \sum_{i' = 1}^{D} \frac{1}{D}  \bm{L}_{\hat{\bm{G}}} \bm{L}_{\bm{V}}^{-1} \lsk \hat{\bm{u}}_{i'} - \frac{1}{D} \sum_{l=1}^{D} \hat{\bm{u}}_{l} \rsk \\
&= \bm{L}_{\hat{\bm{G}}} \bm{L}_{\bm{V}}^{-1} \lbk \lsk \frac{1}{D}\sum_{i' = 1}^{D} \hat{\bm{u}}_{i'} \rsk - \lsk \frac{1}{D} \sum_{l=1}^{D} \hat{\bm{u}}_l \rsk   \rbk = \bm{0}_q, \\
    \mathrm{E}^*(\bm{u}_i^* \bm{u}_i^{*\top}) &= \sum_{i' = 1}^{D} \mathbb{P}( \bm{u}_i^* = \hat{\bm{u}}_{i'}^{sc}) \times \hat{\bm{u}}_{i'}^{sc} \hat{\bm{u}}_{i'}^{sc\top} 
    = \sum_{i' = 1}^{D} \frac{1}{D} \times \hat{\bm{u}}_{i'}^{sc} \hat{\bm{u}}_{i'}^{sc\top} = \sum_{i' = 1}^{D} \frac{1}{D} \bm{L}_{\hat{\bm{G}}} \bm{L}_{\bm{V}}^{-1} \hat{\bm{u}}_{i'}^{c}  \hat{\bm{u}}_{i'}^{c\top} \bm{L}_{\bm{V}}^{-1\top} \bm{L}_{\hat{\bm{G}}}^\top \\
    &=  \bm{L}_{\hat{\bm{G}}} \bm{L}_{\bm{V}}^{-1} \lbk \sum_{i' = 1}^{D} \frac{1}{D}  \hat{\bm{u}}_{i'}^{c}  \hat{\bm{u}}_{i'}^{c\top} \rbk\bm{L}_{\bm{V}}^{-1\top} \bm{L}_{\hat{\bm{G}}}^\top = \bm{L}_{\hat{\bm{G}}} \bm{L}_{\bm{V}}^{-1} \bm{L}_{\bm{V}} \bm{L}_{\bm{V}}^\top \bm{L}_{\bm{V}}^{-1\top} \bm{L}_{\hat{\bm{G}}}^\top =  \bm{L}_{\hat{\bm{G}}} \bm{L}_{\hat{\bm{G}}}^\top = \hat{\bm{G}}.
\end{align*}
Moreover, the bootstrap samples for the decorrelated residuals $w_{ij}^*$ are obtained by first sampling a donor cluster $d_i^*$ using PPS sampling and then by a SRS within the vector of decorrelated residuals $(\hat{w}_{d_i^* 1}^{sc},\cdots, \hat{w}_{d_i^* n_{d_i^*}}^{sc})^\top$. Therefore, we have
\begin{align*}  
    \mathrm{E}^*(w_{ij}^*) &= \sum_{i' = 1}^{D} \sum_{j' = 1}^{n_{i'}} \mathbb{P}(d_i^* = i') \mathbb{P}(w_{ij}^* = \hat{w}_{i'j'}^{sc}) \times \hat{w}_{i'j'}^{sc} 
= \sum_{i' = 1}^{D} \sum_{j' = 1}^{n_{i'}} \frac{n_{i'}}{N} \frac{1}{n_{i'}}  \times \hat{w}_{i'j'}^{sc} \\
&= \frac{\hat{\sigma}_e \sqrt{1 - \hat{\rho}^2} }{\sqrt{N^{-1} \sum_{l = 1}^{D} \sum_{m = 1}^{n_{l}} (\hat{w}_{lm}^c)^2  }} \sum_{i' = 1}^{D} \frac{n_{i'}}{N} \frac{1}{n_{i'}} \sum_{j' = 1}^{n_{i'}}  \hat{w}_{i'j'}^{c} = 0 \\
&= \frac{\hat{\sigma}_e \sqrt{1 - \hat{\rho}^2} }{\sqrt{N^{-1} \sum_{l = 1}^{D} \sum_{m = 1}^{n_{l}} (\hat{w}_{lm}^c)^2  }} \sum_{i' = 1}^{D} \frac{n_{i'}}{N}
 \lsk \frac{1}{n_{i'}} \sum_{j' = 1}^{n_{i'}} \hat{w}_{i'j'} - \frac{1}{n_{i'}} \sum_{m=1}^{n_{i'}} \hat{w}_{i'm}   \rsk = 0, \\
 \mathrm{E}(w_{ij}^{*2}) &= \sum_{i' = 1}^{D} \sum_{j' = 1}^{n_{i'}} \mathbb{P}(d_i^* = i') \mathbb{P}(w_{ij}^* = \hat{w}_{i'j'}^{sc}) \times (\hat{w}_{i'j'}^{sc})^2 = \sum_{i' = 1}^{D} \sum_{j' = 1}^{n_{i'}} \frac{n_{i'}}{N} \frac{1}{n_{i'}}  \times (\hat{w}_{i'j'}^{sc})^2\\
&=\frac{\hat{\sigma}_e^2 (1 - \hat{\rho}^2)}{N^{-1} \sum_{l = 1}^{D} \sum_{m = 1}^{n_{l}} (\hat{w}_{lm}^c)^2  } \sum_{i' = 1}^{D} \frac{n_{i'}}{N} \frac{1}{n_{i'}} \sum_{j' = 1}^{n_{i'}} ( \hat{w}_{i'j'}^c)^2 \\
&= \frac{\hat{\sigma}_e^2 (1-\hat{\rho}^2) }{N^{-1} \sum_{l = 1}^{D} \sum_{m = 1}^{n_{l}} (\hat{w}_{lm}^c)^2  } N^{-1}\sum_{i' = 1}^{D} \sum_{j' = 1}^{n_{i'}}  (\hat{w}_{i'j'}^c)^2 = \hat{\sigma}_e^2 (1-\hat{\rho}^2).
\end{align*}

Recall that the bootstrap samples for the residuals are constructed as $e_{ij}^* = \hat{\rho} e_{i,j-1}^* + w_{ij}^*$ for $j=2, \cdots, n_i$ and $e_{i1}^* = w_{i1}^* / \sqrt{1 - \hat{\rho}^2}$, which is equivalent to the representation 
\begin{equation*}
    e_{ij}^* = \hat{\rho}^{j-1} \frac{w_{i1}^*}{\sqrt{1 - \hat{\rho}^2}} + \sum_{m = 2}^{j} \hat{\rho}^{j-k} w_{im}^*.
\end{equation*}
Then, we obtain
\begin{align*}
    \mathrm{E}(e_{ij}^*) &= \hat{\rho}^{j-1} \frac{ \mathrm{E} (w_{i1}^*)}{\sqrt{1 - \hat{\rho}^2}} + \sum_{m = 2}^{j} \hat{\rho}^{j-k}  \mathrm{E}(w_{im}^*) = 0, \\
    \mathrm{E}(e_{ij}^* e_{ij'}^*) 
    &= \mathrm{E}\lbk \lsk \hat{\rho}^{j-1} \frac{w_{i1}^*}{\sqrt{1 - \hat{\rho}^2}} + \sum_{m = 2}^{j} \hat{\rho}^{j-k} w_{im}^*\rsk \lsk \hat{\rho}^{j'-1} \frac{w_{i1}^*}{\sqrt{1 - \hat{\rho}^2}} + \sum_{m = 2}^{j'} \hat{\rho}^{j'-k} w_{im}^* \rsk \rbk \\
    &= \hat{\rho}^{j-1} \hat{\rho}^{j'-1} \frac{ \mathrm{E}(w_{i1}^{*2}) }{1 - \hat{\rho}^2} + \sum_{m=2}^{\min(j,j')} \hat{\rho}^{j-m} \hat{\rho}^{j' - m} \mathrm{E}(w_{im}^{*2}) \\
    &= \hat{\rho}^{j + j' -2} \frac{ \hat{\sigma}^2_e (1 - \hat{\rho}^2)  }{1 - \hat{\rho}^2} + \hat{\sigma}^2_e (1 - \hat{\rho}^2) \rho^{j + j' - 2\min(j,j')} \sum_{m=2}^{\min(j,j')} \hat{\rho}^{2(\min(j-j') - m)} \\
    &= \hat{\rho}^{j + j' -2} \hat{\sigma}^2_e + \hat{\sigma}^2_e (1 - \hat{\rho}^2) \hat{\rho}^{j + j' - 2\min(j,j')} \sum_{m=0}^{\min(j,j') - 2} \hat{\rho}^{2m} \\
    &= \hat{\rho}^{j + j' -2} \hat{\sigma}^2_e + \hat{\sigma}^2_e (1 - \hat{\rho}^2) \hat{\rho}^{j + j' - 2\min(j,j')} \frac{ 1 - \hat{\rho}^{2(\min(j,j') - 1)}  }{1 - \hat{\rho}^2} \\
    &= \hat{\rho}^{j + j' -2} \hat{\sigma}^2_e + \hat{\sigma}^2_e \hat{\rho}^{j + j' - 2\min(j,j')} - \hat{\sigma}^2_e \hat{\rho}^{j + j' -2} \\
    &= \hat{\sigma}^2_e \hat{\rho}^{j + j' - 2\min(j,j')} = \hat{\sigma}^2_e \hat{\rho}^{|j-j'|}.
\end{align*}

Therefore, the PREB bootstrap satisfies the consistency conditions, and thus its corresponding bootstrap percentile confidence intervals are consistent according to \cite{shaoETAL2000} and \cite{carpenterETAL2003}. Importantly, this consistency holds under very general clustered data settings, including imbalanced cluster sizes $n_i$, the presence of random slopes in $\bm{u}_i$ (i.e., $q > 1$), and within-cluster autocorrelation captured by $\bmSigma(\bm{G}, \sigma^2_e, \rho)$, owing to the combined use of reflation step in \eqref{eq:hatu_hatepsilon_PREB}, PPS sampling for selecting donor clusters $d_i^*$, and reconstruction of AR-1 bootstrap samples $e_{ij}^*$ from decorrelated residuals $w_{ij}^*$. Moreover, since the quasi-ML estimating functions in \eqref{eq:score} are used to obtain the estimator $\hat{\bmtheta}$ under the quasi-Gaussian likelihood, the preceding theoretical results are valid provided that the first two moments of the random effects and error terms in model \eqref{eq:lmm} are correctly specified.  We do not require their full distributions so the results hold even when the random effect and error terms are not Gaussian. In contrast, it can be shown that the original prescaled REB bootstrap of \cite{chambersANDchandra2013} satisfies the consistency conditions only under the restrictive setting of $n_i = n$ for $i=1,\cdots,D$ (balanced cluster sizes), $q = 1$ (random-intercept only), and $\rho = 0$ (independent error terms).

\section{Simulation Study} \label{sec:simulation}
We conducted a simulation study to evaluate the finite sample inferential performance of the proposed PREB bootstrap for clustered data. The true data generating process followed \eqref{eq:lmm} with $p=2$, $\bmbeta = (1,2)^\top$, $\bm{x}_{ij} = (1,x_{ij,1})^\top$, and $x_{ij,1} \sim U(0,1)$. The covariates $\bm{x}_{ij}$ were fixed across simulation replicates.  

We considered two settings for the dimension of the random effect vector $q$ (Sets 1 -- 2), together with four scenarios for generating the random effects and error terms (Sets A -- D). Specifically, in Set 1 we set $q = 2$ with $\bm{z}_{ij} = \bm{x}_{ij}$ and 
\begin{equation*}
    \bm{G} = \begin{pmatrix}
        0.04 & 0.02 \\
        0.02 & 0.04
    \end{pmatrix}
\end{equation*}
corresponding to a random intercept and random slope model. In Set 2, we set $q = 1$ with $\bm{z}_{ij} = 1$ and $\bm{G} = 0.04$, corresponding to a random intercept model. For the generation of $\bm{u}_i$ and $e_{ij}$, we fixed $\sigma^2_e = 0.16$ and considered the following four settings:
\begin{itemize}
    \item \textbf{Set A}: $\bm{u}_i \stackrel{i.i.d.}{\sim} N(\bm{0}_q, \bm{G})$, $e_{ij} \stackrel{i.i.d.}{\sim} N(0, \sigma^2_e)$.
    \item \textbf{Set B}: $\bm{u}_i = \bm{L}_{\bm{G}} (\eta_{i0},\cdots,\eta_{i,q-1})^\top$ where $\eta_{ik} \stackrel{i.i.d.}{\sim} (\chi^2_1 - 1)/\sqrt{2}$ for $k=0,\cdots,q-1$ with $\bm{G} = \bm{L}_{\bm{G}} \bm{L}_{\bm{G}}^\top$ and $\chi^2_1$ denotes a chi-squared random variable with one degree of freedom, $e_{ij} \stackrel{i.i.d.}{\sim} \sigma_e \{ (\chi^2_1 - 1)/\sqrt{2} \}$ .
    \item \textbf{Set C:} $\bm{u}_i \stackrel{i.i.d.}{\sim} N(\bm{0}_q, \bm{G})$, and $e_{ij} = \rho e_{i,j-1} + w_{ij}$ for $j=2,\cdots,n_i$, where $\rho = 0.5$, $w_{ij} \stackrel{i.i.d.}{\sim} N(0,\sigma_e^2 (1-\rho^2))$ and $e_{i1} \stackrel{i.i.d.}{\sim} N(0, \sigma^2_e)$.
    \item \textbf{Set D}: $\bm{u}_i \stackrel{i.i.d.}{\sim} N(\bm{0}_q, \bm{G})$, and $e_{ij} = \rho e_{i,j-1} + w_{ij}$ for $j = 2,\cdots,n_i$ and $i=1,\cdots,D$, together with $e_{i1} = \rho e_{i-1,n_{i - 1} } +  w_{i1}$ for $i=2,\cdots,D$, where $\rho = 0.5$, $w_{ij} \stackrel{i.i.d.}{\sim} N(0,\sigma_e^2 (1-\rho^2))$ and $e_{11} \sim N(0, \sigma^2_e)$.
\end{itemize}
Thus, Set A generates Gaussian random effects and independent Gaussian error terms, while Set B considers non-Gaussian random effects and errors based on the centered and standardized chi-squared distribution. Set C introduces within-cluster AR-1 autocorrelation in the error terms, whereas Set D  further induces dependence between clusters by generating all $N = \sum_{i=1}^{D} n_i$ errors from a single AR-1 process. Sets A -- D are similar to those considered in the simulation study of \cite{chambersANDchandra2013} who focused on balanced cluster sizes and random-intercept-only model ($q=1$).

\begin{figure}[!tb]
    \centering
    \includegraphics[width=0.7\linewidth]{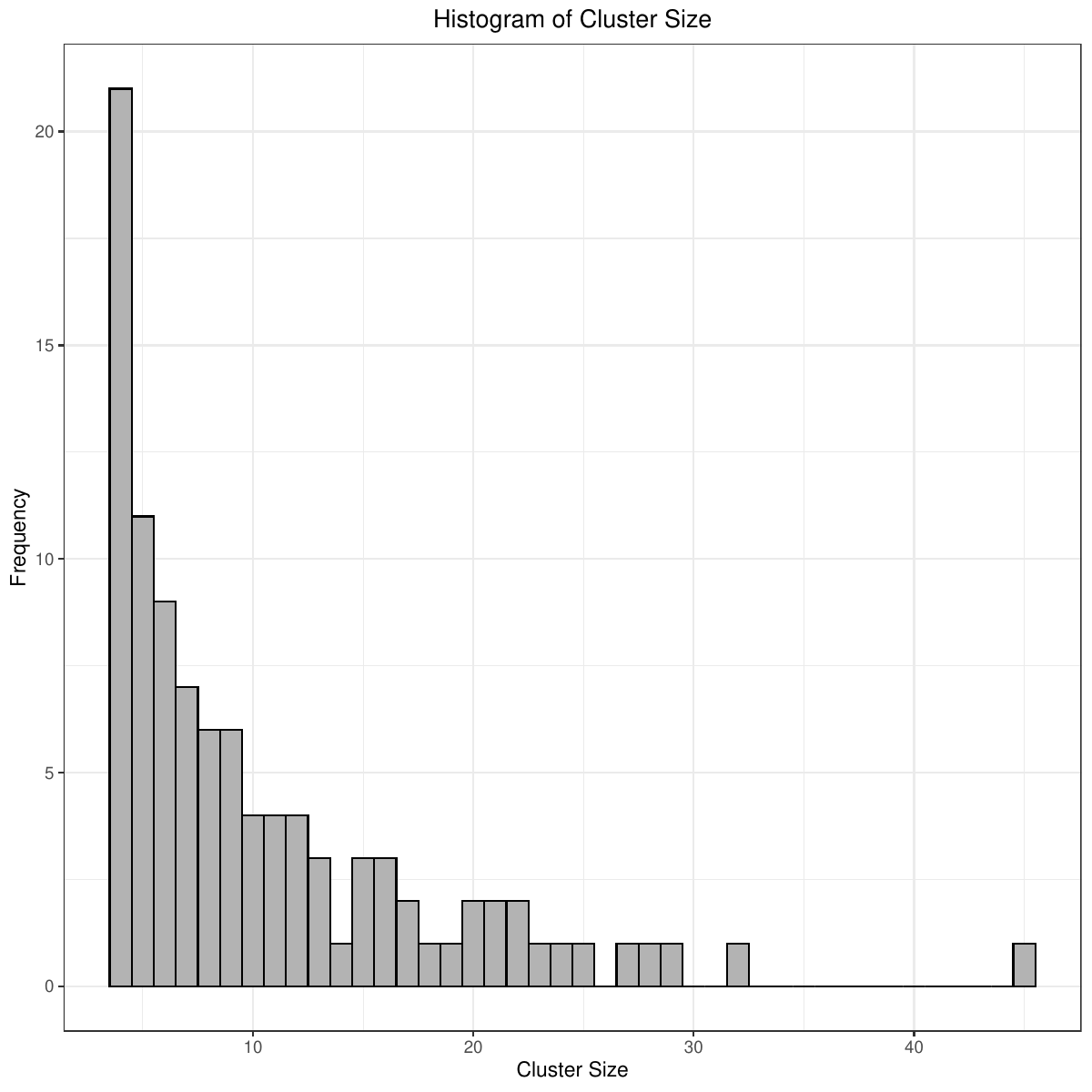}
    \caption{Histogram of imbalanced cluster sizes $\{\tilde{n}_i: i=1,\cdots,100\}$ in the simulation.}
    \label{fig:hist_ni_sim}
\end{figure}

We fixed the number of clusters to be $D = 100$ and considered imbalanced cluster sizes as shown in Figure \ref{fig:hist_ni_sim}, with cluster sizes ranging from 4 to 45. Denote this set of cluster sizes as $\{\tilde{n}_i: i = 1,\cdots,D\}$. For Set 1 with random slopes, we considered three cluster size configurations: $n_i = \tilde{n}_i$, $n_i = \tilde{n}_i + 1$, and $n_i = 10$, corresponding to imbalanced design with $\min_i(n_i) = 4$, imbalanced design with $\min_i(n_i) = 5$, and a balanced design, respectively. For Set 2 with a random intercept only, we instead considered $n_i = \tilde{n}_i - 3$, $n_i = \tilde{n}_i - 2$, and $n_i = 10$, corresponding to imbalanced design with $\min_i(n_i) = 1$, imbalanced design with $\min_i(n_i) = 2$, and a balanced design, respectively. Different cluster size settings were used for Sets 1 and 2 because fitting LMMs with $q = 2$ tends to become numerically unstable when the minimum cluster size is very small. Consequently, for Set 1 we only considered settings with $\min_i(n_i) \geq 4$.  The coefficient of variation $ \{ D^{-1} \sum_{i=1}^{D} (n_i - D^{-1} \sum_{i' = 1}^{D} n_{i'})^2  \}^{1/2} / (D^{-1} \sum_{i=1}^{D} n_i) $ equals 1.007, 0.889, 0.720, 0.657, and 0 for the imbalanced designs with $\min_i(n_i) = 1,2,4,5$ and the balanced design, respectively. This allows us to investigate how bootstrap performance changes as the degree of imbalance decreases (as $\min_i(n_i)$ increases) and approaches the balanced case. For each of the 24 settings (2 settings for $q$ $\times$ 4 settings for the generation of $\bm{u}_i$ and $e_{ij}$ $\times$ 3 cluster size configurations), we generated a total of 100 datasets. In a few cases under Set 1 (random slope model), datasets were discarded due to a singular $\hat{\bm{G}}$, which prevented computation of $\bm{L}_{\hat{\bm{G}}}$; these were replaced to maintain 100 valid datasets per setting.

We applied the proposed PREB bootstrap to each simulated dataset to construct $95\%$ bootstrap percentile confidence intervals for $\bmtheta = (\bmbeta^\top, \mathrm{vech}(\bm{G})^\top, \sigma^2_e, \rho)^\top$. For comparison, we implemented several alternative bootstrap methods for clustered data:
\begin{itemize}
    \item \textbf{Prescaled REB bootstrap} \citep{chambersANDchandra2013}, extended to $q > 1$ by using the Cholesky factor of $D^{-1}\sum_{i=1}^{D} \hat{\bm{u}}_i \hat{\bm{u}}_i^\top$ in place of $\bm{L}_{\bm{V}}$ in \eqref{eq:hatu_hatepsilon_PREB} for reflation, using SRS instead of PPS sampling for donor cluster sampling, and omitting the decorrelation step for residuals.
    \item \textbf{REB-nc} \citep{relugaETAL2024}, a modification of the above without donor cluster resampling (i.e., $d_i^* = i$).
    \item \textbf{CGR bootstrap} \cite{carpenterETAL2003}, which replaces $\hat{\bm{u}}_i$ with the EBLUPs $\tilde{\bm{u}}_i$, and samples the residuals $\bm{e}^* = \mathrm{SRSWR}( \{ (N^{-1}\sum_{i=1}^{D} \sum_{j=1}^{n_i}  \tilde{e}_{ij}^2 )^{-1/2} \tilde{e}_{11}, \cdots,  (N^{-1}\sum_{i=1}^{D} \sum_{j=1}^{n_i}  \tilde{e}_{ij}^2 )^{-1/2} \\ \tilde{e}_{Dn_D}  \}, N)$, where $\tilde{e}_{ij} = y_{ij} - \bm{x}_{ij}^\top \hat{\bmbeta} - \bm{z}_{ij}^\top\tilde{\bm{u}}_i$ are EBLUP-based residuals.
    \item \textbf{Cluster bootstrap} \citep{davisonANDhinkley1997, mccullagh2000}, which resamples entire clusters, i.e., $\bm{y}^* = (\bm{y}_{h_1^*}^\top,\cdots, \bm{y}_{h_D^*}^\top)^\top$, $\bm{X}^* = (\bm{X}_{h_1^*}^\top, \cdots, \bm{X}_{h_D^*}^\top)^\top$, and $\bm{Z}^* = (\bm{Z}_{h_1^*}^\top, \cdots, \bm{Z}_{h_D^*}^\top)^\top$ with $h_i^* = \mathrm{SRSWR}(\{1,\cdots,D\},1)$, and fits model \eqref{eq:lmm} to $(\bm{y}^*, \bm{X}^*, \bm{Z}^*)$ assuming independent errors.
    \item \textbf{Cluster AR-1 bootstrap}, a variant of the cluster bootstrap by fitting model \eqref{eq:lmm} to $(\bm{y}^*, \bm{X}^*, \bm{Z}^*)$ assuming within-cluster AR-1 error structure.
    \item \textbf{Parametric bootstrap}, which generates $\bm{u}_i^* \stackrel{i.i.d.}{\sim} N(\bm{0}_q, \hat{\bm{G}})$ and $e_{ij}^* \stackrel{i.i.d.}{\sim} N(0, \hat{\sigma}^2_e)$.
\end{itemize}
For PREB and cluster AR-1 bootstraps, the bootstrapped datasets were used to fit model \eqref{eq:lmm} with an AR-1 error structure within each cluster. For the remaining methods, the bootstrapped datasets were used to fit model \eqref{eq:lmm} with i.i.d. errors, which is consistent with their generation of bootstrapped errors $e_{ij}^*$. Similarly, for the original simulated data, a LMM with AR-1 error was fitted for the PREB bootstrap, while a LMM assuming independent errors was fitted for REB, REB-nc, CGR and parametric bootstraps to obtain the initial estimators. 

For all methods, $B = 500$ bootstrap replicates were used. We assessed performance based on the empirical coverage rates of 95\% bootstrap percentile confidence intervals for $\bmtheta$, 
computed over the 500 simulated datasets. We also considered the basic bootstrap confidence interval \citep{davisonANDhinkley1997}, which constructs the 95\% confidence interval for the $k$-th element of $\bmtheta$ as $(2 \hat{\theta}_k - \hat{\theta}_{k, 0.975}^*, 2 \hat{\theta}_k -  \hat{\theta}_{k, 0.025}^*)$, where $\hat{\theta}_{k,p}^*$ denote the $p$-th quantile of the bootstrap distribution for $\hat{\theta}_k$, and $\hat{\theta}_k$ is the $k$-th component of the original estimate $\hat{\bmtheta}$. The results based on the basic interval were very similar to those from the percentile interval; therefore, for brevity, we only report results for the percentile interval below. Results for the basic interval are available from the authors upon request.

\begin{figure}[!tb]
    \centering
    \includegraphics[width=1\linewidth]{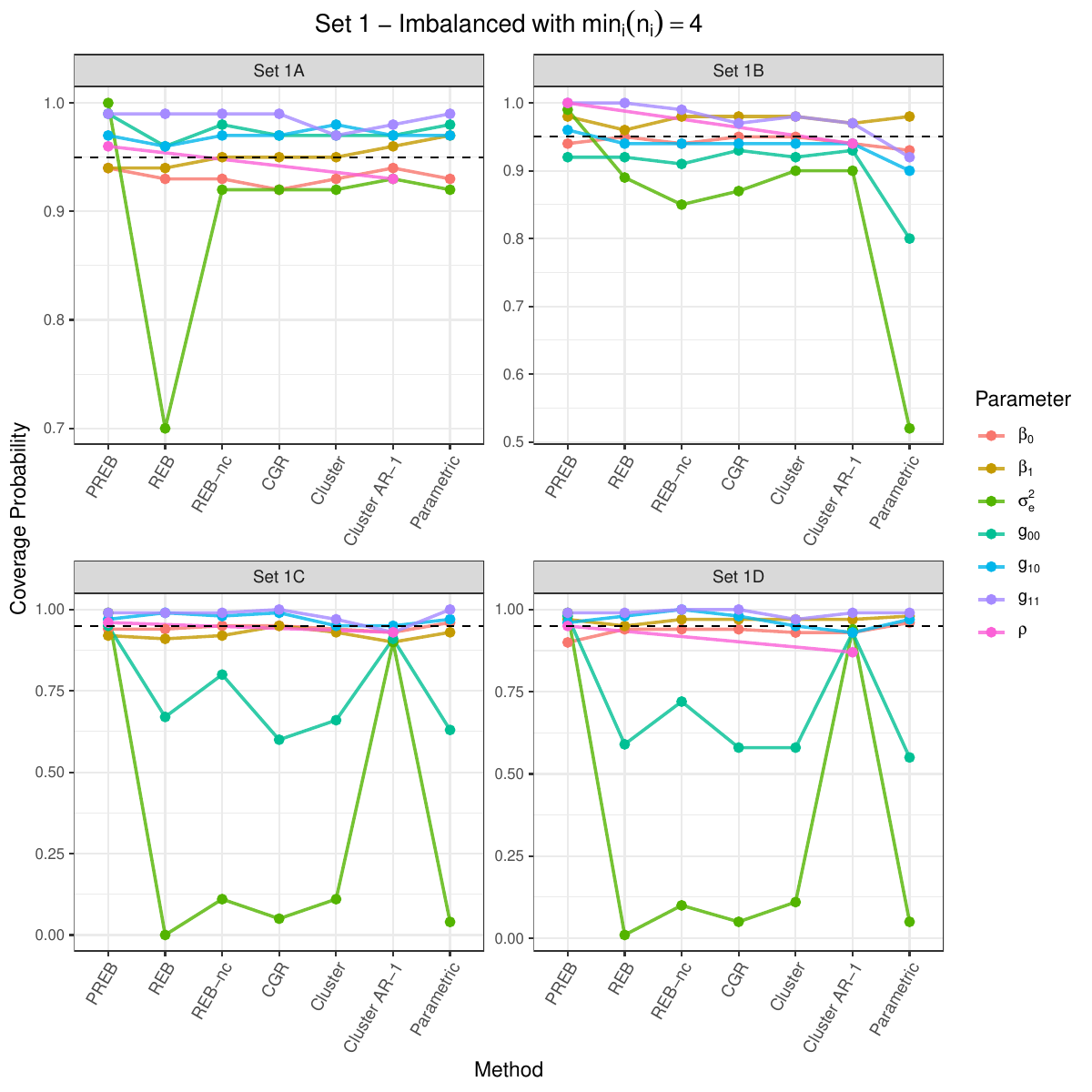}
    \caption{Empirical coverage rates (y-axis) of 95\% percentile confidence intervals across bootstrap methods (x-axis) for different parameters (indicated by colors) under Set 1A -- 1D (panels), for the imbalanced setting with $\min_i(n_i) = 4$. }
    \label{fig:Set1_min_ni_4}
\end{figure}

\begin{figure}[!tb]
    \centering
    \includegraphics[width=1\linewidth]{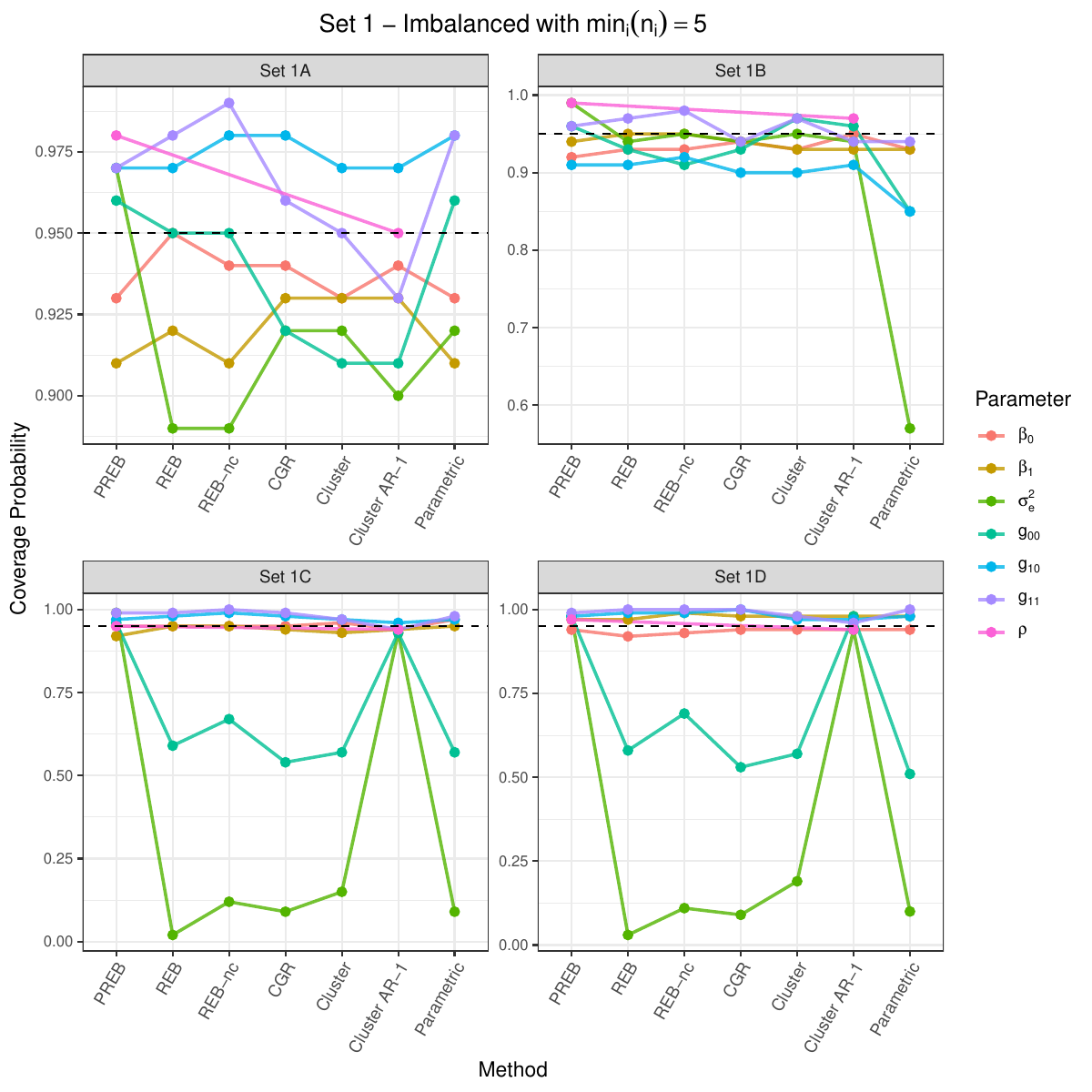}
    \caption{Empirical coverage rates (y-axis) of 95\% percentile confidence intervals across bootstrap methods (x-axis) for different parameters (indicated by colors) under Set 1A -- 1D (panels), for the imbalanced setting with $\min_i(n_i) = 5$. }
    \label{fig:Set1_min_ni_5}
\end{figure}

\begin{figure}[!tb]
    \centering
    \includegraphics[width=1\linewidth]{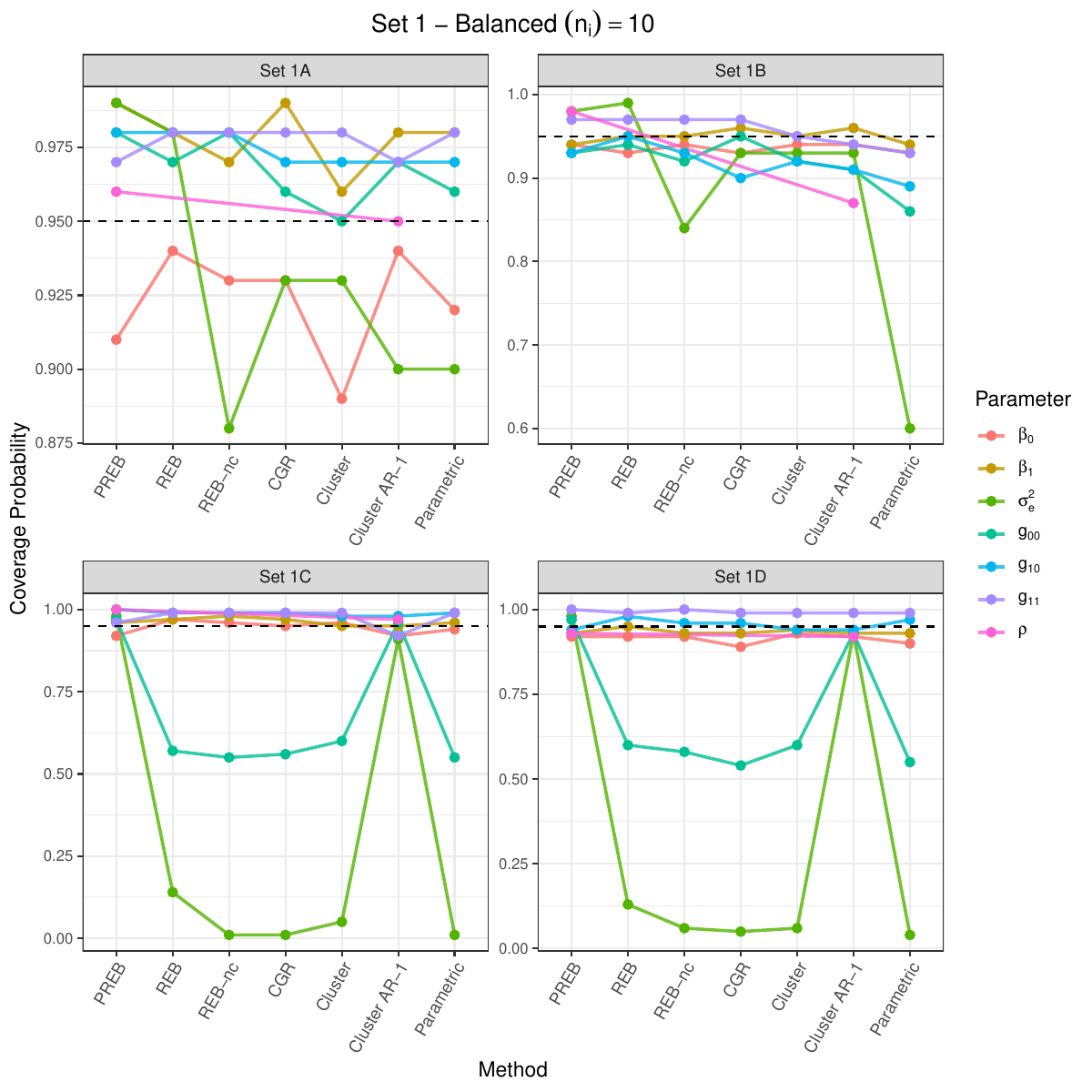}
    \caption{Empirical coverage rates (y-axis) of 95\% percentile confidence intervals across bootstrap methods (x-axis) for different parameters (indicated by colors) under Set 1A -- 1D (panels), for the balanced setting with $n_i = 10$. }
    \label{fig:Set1_bal_ni10}
\end{figure}

Figures \ref{fig:Set1_min_ni_4} -- \ref{fig:Set1_bal_ni10} present the results for Sets 1A -- 1D under the three cluster size configurations. The proposed PREB bootstrap consistently achieves coverage rates close to the nominal 95\% level for all parameters in $\bmtheta$, demonstrating its ability to accommodate a wide range of clustered data settings, including the Gaussian case with independent errors (Set A), non-normal random effects and error terms (Set 1B), within-cluster AR-1 errors (Set 1C), and AR-1 errors extending across clusters (Set 1D). Moreover, the stable performance of the PREB bootstrap across all cluster size settings confirms its applicability under both balanced and imbalanced designs with varying degrees of imbalance. Since Set 1 corresponds to the random slope model with $q=2$, these results also demonstrate its ability to handle both random intercepts and random slopes. 

Compared with the competing bootstrap methods, the advantages of the PREB bootstrap are most pronounced under Sets 1C and 1D, where AR-1 error dependence is present. In these settings, the proposed bootstrap substantially outperforms alternative methods (except cluster AR-1 bootstrap) that often exhibit substantial undercoverage for $g_{00}$ and $\sigma^2_e$. This is unsurprising because these methods implicitly assume independence among the error terms in their bootstrap procedure e.g., the CGR bootstrap generates bootstrap residuals by SRS from the pooled $N$-dimensional vector of EBLUP-based residuals, effectively treating all $N$ residuals as independent.

Comparing the proposed bootstrap to the original REB bootstrap, the REB bootstrap exhibits noticeable undercoverage for $\sigma^2_e$ under the most imbalanced setting with $\min_i(n_i) = 4$ in Sets 1A and 1B. As the degree of imbalance decreases in the setting with $\min_i(n_i) = 5$, the REB coverage for $\sigma^2_e$ improves, with only slight undercoverage remaining in Set 1A. Under the balanced design, it achieves coverage rates close to 95\% for all parameters in both Sets 1A and 1B. This progressive improvement from the most imbalanced to balanced setting is consistent with the original design of REB bootstrap, which was primarily developed for balanced clustered data. 
These comparisons, including those in Sets 1C and 1D, therefore suggest that the PREB bootstrap has successfully extends the REB framework to accommodate imbalanced clusters, random slopes, and autocorrelated errors.

The parametric bootstrap performs poorly under Set 1B, showing undercoverage for $\sigma^2_e$, $g_{00}$ and $g_{10}$ (covariance between random intercept and random slope), reflecting its sensitivity to the Gaussian assumptions for the random effects and error terms. Similarly, the REB-nc bootstrap also tends to under-cover $\sigma^2_e$ under Set 1B. The two cluster bootstrap variants perform reasonably well under Sets 1A and 1B across all cluster size configurations. However, under Sets 1C and 1D, the cluster AR-1 bootstrap continues to maintain good coverage, whereas the standard cluster bootstrap exhibits undercoverage for $\sigma^2_e$ and $g_{00}$. This highlights the importance of incorporating AR-1 error structure into the fitted LMM when the underlying errors are autocorrelated. An important caveat for the two variants of the cluster bootstrap under imbalanced $n_i$ settings is that the resulting bootstrapped datasets $(\bm{y}^*, \bm{X}^*, \bm{Z}^*)$ do not necessarily preserve the original total sample size $N$, whereas the proposed PREB bootstrap does.

The results for Sets 2A -- 2D under the three cluster size configurations are provided in the supplementary material, as they are qualitatively similar to those for Sets 1A -- 1D. Overall, the proposed bootstrap continues to perform well under Set 2 (with $q = 1$) across various clustered data structures, and generally outperforms the competing bootstrap methods, particularly in terms of coverage for $\sigma^2_e$ and $g_{00}$ in Sets 2C and 2D. One additional finding under Set 2 is that most alternative bootstrap methods exhibit undercoverage for $\sigma^2_e$ and $g_{00}$ in Set 2B, whereas the PREB bootstrap continues to maintain coverage rates close to the nominal level.


\section{Application to Mayo Clinic Primary Biliary Cirrhosis Dataset} \label{sec:realdata}

\begin{figure}[!tb]
    \centering
    \includegraphics[width=0.7\linewidth]{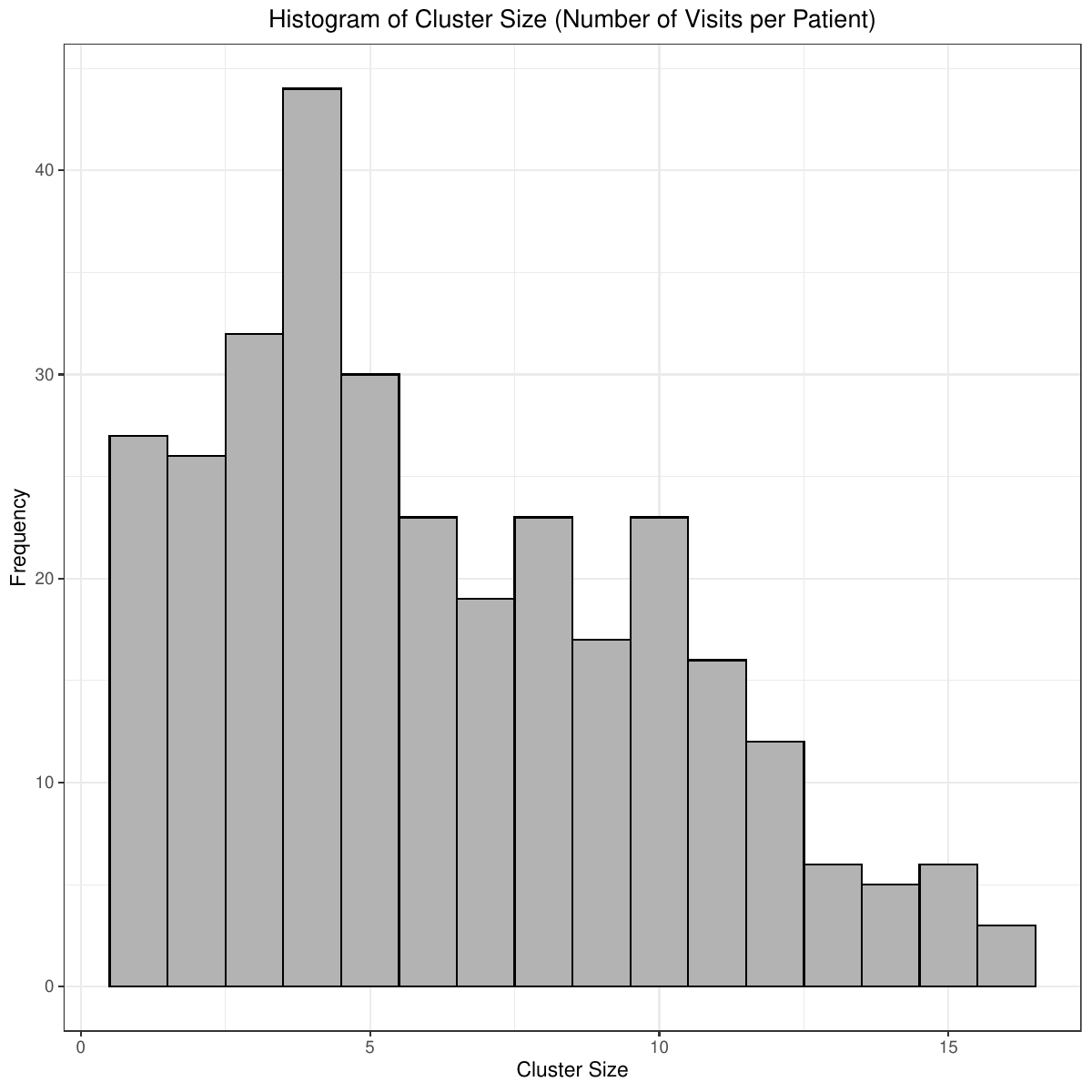}
    \caption{Histogram of cluster sizes $n_i$ for $i=1,\cdots,312$ patients in the Mayo Clinic primary biliary cirrhosis data.}
    \label{fig:realdata_hist_ni}
\end{figure}

We apply the PREB bootstap to a longitudinal study of the primary biliary cirrhosis (PBC) disease conducted by the Mayo Clinic. The dataset, which is available in the \texttt{survival} \texttt{R} package, contains repeated biomarker measurements for $D =312$ PBC patients recruited at the Mayo Clinic between 1974 and 1984. These patients participated in two randomized clinical trials evaluating D-penicillamine for the treatment of PBC and were followed until 1988, with most attending annual follow-up visits and resulting in a total of $N = 1945$ total number of observations. The repeated visits are nested within patients giving rise to a clustered data structure. Figure \ref{fig:realdata_hist_ni} shows the distribution of the number of visits per patient (cluster sizes $n_i$), indicating cluster size imbalance with sizes ranging from 1 to 16. 

At each visit, multiple biochemical and clinical variables were recorded, yielding a longitudinal dataset consisting of two demographic variables (age and sex), a time-varying covariate (time of visit), seven categorical variables (including treatment status, end-of-trial status, hepatomegaly, and ascites indicators), and six continuous biomarkers (serum bilirubin, serum albumin, alkaline phosphatase, aspartate aminotransferase, standardized blood clotting time, and platelet count). This dataset has been widely used in the survival analysis literature \citep{markusETAL1989,flemingANDthomas1991,murtaughETAL1994} to study survival modelling for PBC patients. 

More recently, \cite{wang2017} and \cite{taavoniANDarashi2022} studied the joint modelling of log-transformed \texttt{bili} (serum bilirubin) and log-transformed serum albumin under the framework of multivariate $t$ linear mixed models. Given that \texttt{bili} is a key indicator of liver disease and that elevated levels of \texttt{bili} excreted in bile and urine may indicate disease, we follow these studies and model log-transformed \texttt{bili} as the response variable $y_{ij}$ (with $i$ indexing patients and $j$ indexing visits) in model \eqref{eq:lmm}. We include all variables described above as fixed effects, resulting in $p = 20$, with the other five continuous biomarker covariates also log-transformed. As in \cite{wang2017}, we consider a random-intercept model ($q = 1$) to capture the between-patient heterogeneity in baseline \texttt{bili}, and estimate the model using quasi-REML.

\begin{figure}[!tb]
    \centering
    \includegraphics[width=0.7\linewidth]{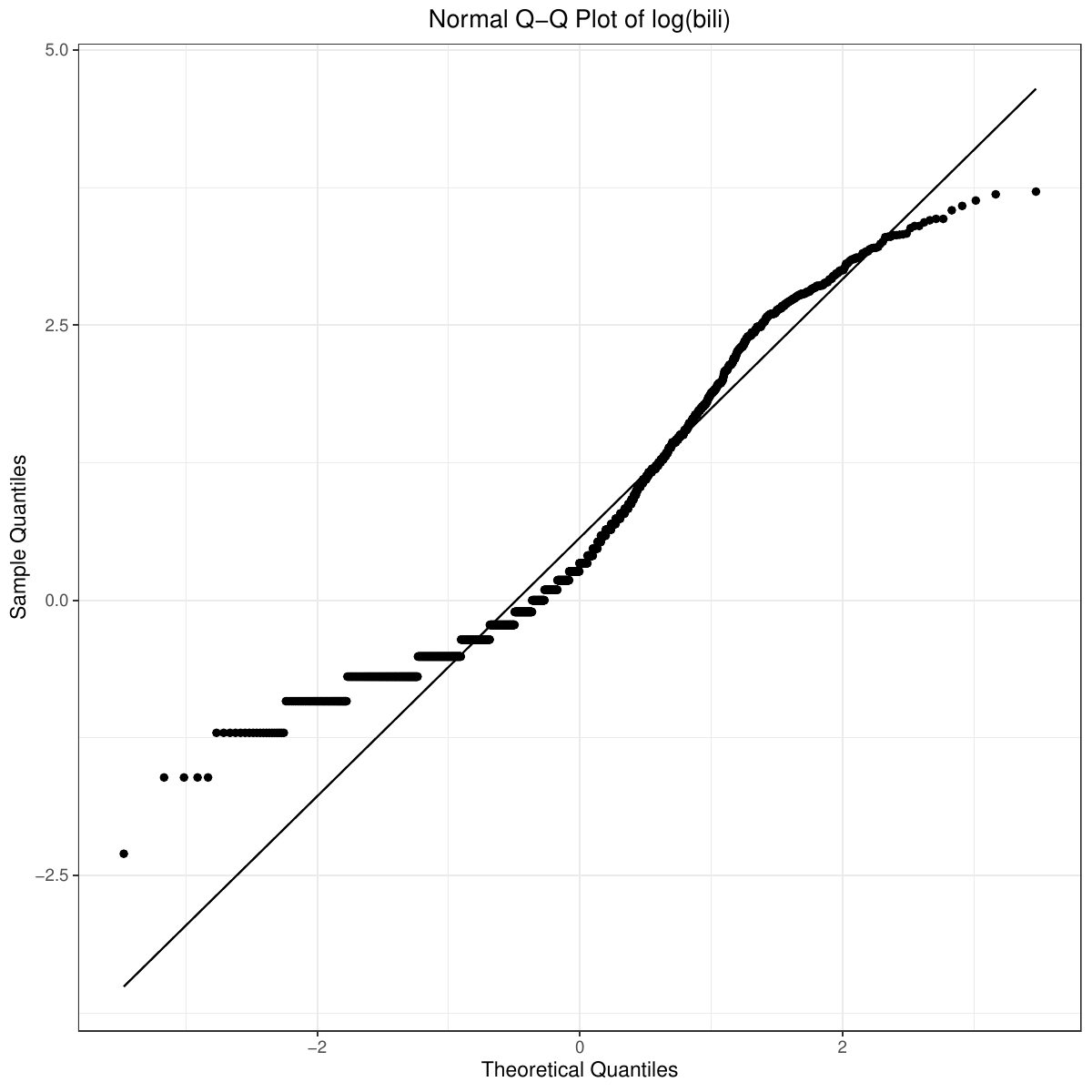}
    \caption{QQplot of log-transformed \texttt{bili} in the Mayo Clinic primary biliary cirrhosis data.}
    \label{fig:realdata_qq_bili}
\end{figure}

Figure \ref{fig:realdata_qq_bili} shows that, even after log-transformation, $y_{ij}$ does not appear to follow a normal distribution. Moreover, due to the longitudinal nature of the data, within-patient autocorrelation in log-\texttt{bili} is expected. This is consistent with the findings of \cite{wang2017} and \cite{taavoniANDarashi2022}, who reported serial dependence across visits for these biomarkers. The combination of substantial imbalance in number of visits, within-patient autocorrelation, and non-normality provides strong motivation to apply the PREB bootstrap for inferences of the LMM parameters. For comparison, we also apply the alternative bootstrap methods considered in Section \ref{sec:simulation} to the same dataset, with $B = 500$ bootstrap replicates for all methods.

\begin{figure}[!tb]
    \centering
    \includegraphics[width=1\linewidth]{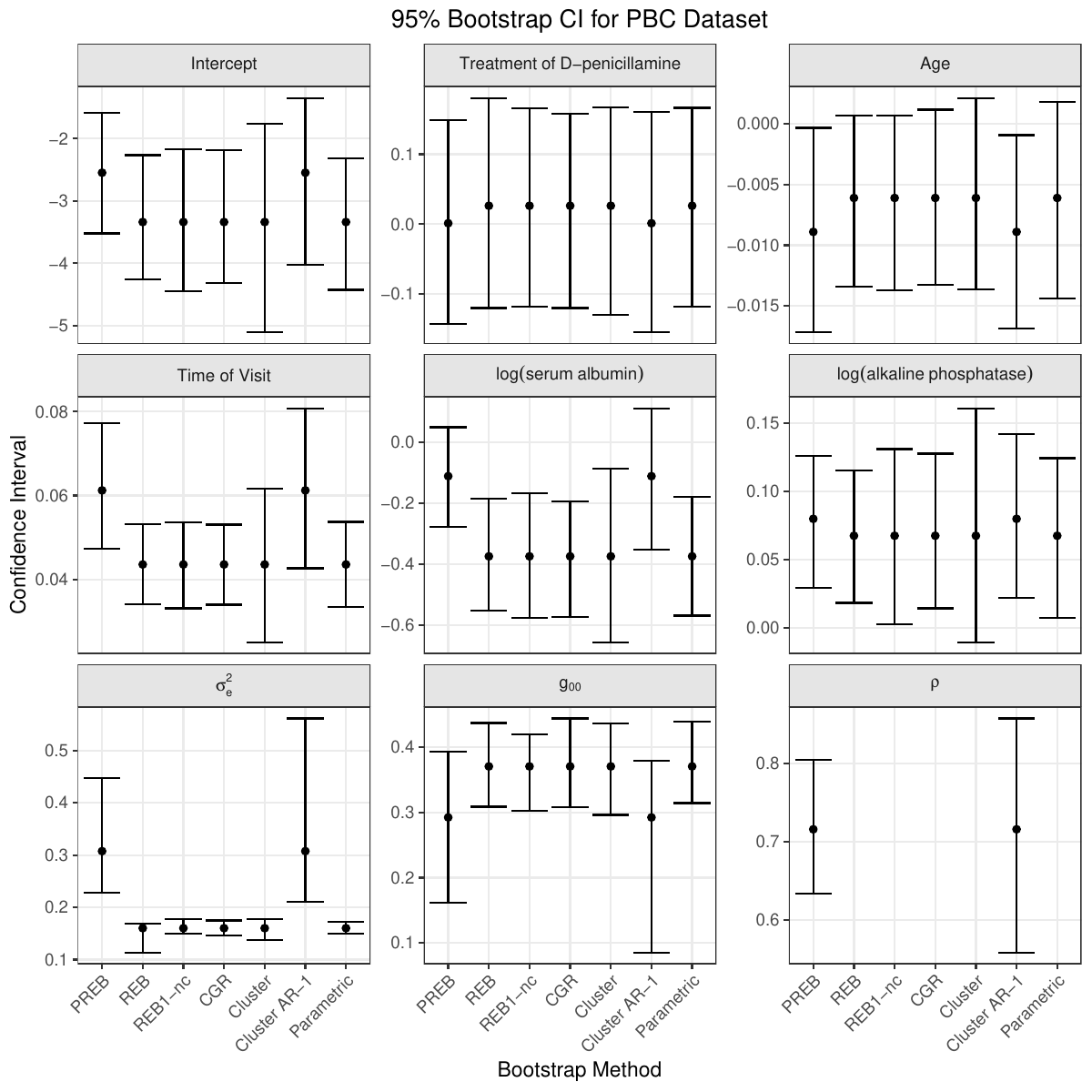}
    \caption{95\% percentile confidence intervals across bootstrap methods (x-axis) for different parameters (panels), together with the corresponding point estimates (indicated by points), for the Mayo Clinic primary biliary cirrhosis data. Results are presented for a subset of the fixed-effect coefficients; see the supplementary material for full results.}
    \label{fig:realdata_subset}
\end{figure}

Figure \ref{fig:realdata_subset} presents the 95\% bootstrap percentile confidence intervals (CIs) for a subset of $\bmbeta$ (full results are given in the supplementary material), as well as for $\sigma^2_e$, $g_{00}$, and $\rho$, together with the corresponding point estimates obtained from fitting LMMs with AR-1 or i.i.d. error structures, depending on the method. 
The PREB and cluster AR-1 results provide clear evidence of within-patient AR-1 residual autocorrelation, with the 95\% CIs for $\rho$ ranging approximately from 0.6 to 0.8 and 0.55 to 0.85, respectively. The PREB and cluster AR-1 CIs for $\sigma^2_e$ and $g_{00}$ also differ noticeably from the remaining bootstrap methods. This is consistent with the simulation results in Section \ref{sec:simulation}, where the PREB and cluster AR-1 CIs exhibited substantially improved coverage under within-cluster autocorrelation. In contrast, the parametric bootstrap yields the narrowest CI for $\sigma^2_e$, reflecting its tendency toward undercoverage when normality assumptions are violated, as also observed in the simulation studies. While the PREB CIs are generally consistent with those from the cluster AR-1 bootstrap, the PREB bootstrap is preferred because it preserves the overall sample size $N$ in the bootstrap samples, whereas this property is not guaranteed under the cluster AR-1 bootstrap due to the imbalanced $n_i$.

For the fixed effects, the PREB CIs are fairly consistent with those from other methods. In particular, the treatment effect of D-penicillamine on the level of log-\texttt{bili} is negligible, with all CIs including zero, which agrees with the findings of \cite{wang2017} and \cite{taavoniANDarashi2022}. However, the PREB and cluster AR-1 CIs for the coefficients of time of visit and log-transformed serum albumin show some differences compared to alternative bootstraps. It is also worth noting that age is found to have a significant negative association with log-\texttt{bili} at the 5\% significance level under the PREB and cluster AR-1 CIs, whereas the other methods do not indicate a significant effect.

\section{Conclusion}\label{sec:conclusion}
We develop the PREB bootstrap for general clustered data structures, by extending the REB bootstrap originally designed for balanced cluster sizes, random-intercept-only models, and independent error terms. In particular, the proposed method accommodates imbalanced clusters, both random intercepts and random slopes, within-cluster AR-1 correlation, and non-normal random effects and error terms. We demonstrate Fisher consistency of the PREB bootstrap under very general clustered data settings. A key feature of our approach is the use of a reflation step for the predicted random effects and residuals, together with a decorrelation step prior to reconstructing the AR-1 bootstrap error structure. Simulation studies show that the PREB bootstrap delivers superior finite sample inferential performance compared to alternative bootstrap methods when error terms are autocorrelated. It also greatly outperforms the original REB bootstrap under imbalanced cluster size settings, as well as the parametric bootstrap under scenarios with non-normal random effects and error terms. We apply the PREB bootstrap to the Mayo clinic primary biliary cirrhosis data to model the log-transformed bilirubin level and conduct inference on the corresponding LMM parameters. The results indicate evidence of within-patient autocorrelation; consequently, the PREB confidence intervals differ meaningfully from those obtained using alternative bootstrap methods that do not account for such dependence, consistent with the simulation findings. 

A natural extension of the PREB bootstrap is to accommodate non-continuous responses by replacing the linear mixed model \eqref{eq:lmm} with a generalized linear mixed model \citep{breslowANDclayton1993}, where the conditional mean of the response is modelled as a linear function of the covariates and random effects through a suitable link function. This would broaden applicability to fields involving discrete responses in a clustered data setting, such as ecology where $y_{ij}$ are the presence-absence records or counts of species $j$ in sampling site $i$. Another direction is to extend the method to perform inference on linear combinations of fixed effects and random effects $\bm{k}_i^\top \bmbeta + \bm{l}_i^\top \bm{u}_i$, which are also known as cluster-level mixed effects parameters; see the recent work of \cite{relugaETAL2024}  who considered bootstrap inference for such parameters in random slope models. 
Finally, it would be of interest to extend the PREB bootstrap to other forms of within-cluster dependence beyond the current AR-1 structure, such as spatially correlated error processes when observations correspond to spatial rather than temporal units.

\section{Disclosure Statement}
The authors have no conflicts of interest to declare.

\section{Data Availability Statement}
The source code for this article is available at \url{https://github.com/Zy1225/Proportional-and-Modified-REB}.

\begin{center}

{\large\bf SUPPLEMENTARY MATERIAL}

\end{center}

\begin{description}
\item[Supplementary Results and Details:]
Additional results for simulations and the real data application.
\end{description}

\bibliographystyle{apalike} 
\bibliography{bibliography}

\end{document}